# A very luminous magnetar-powered supernova associated with an ultra-long gamma-ray burst

Jochen Greiner[1,2], Paolo A. Mazzali[3,4], D. Alexander Kann[5], Thomas Krühler[6], Elena Pian[7,8], Simon Prentice[3], Felipe Olivares E.[9], Andrea Rossi[5,7], Sylvio Klose[5], Stefan Taubenberger[4,10], Fabian Knust[1], Paulo M.J. Afonso[11], Chris Ashall[3], Jan Bolmer[1,12], Corentin Delvaux[1], Roland Diehl[1], Jonathan Elliott[1,17], Robert Filgas[13], Johan P.U. Fynbo[14], John F. Graham[1], Ana Nicuesa Guelbenzu[5], Shiho Kobayashi[3], Giorgos Leloudas[14,15], Sandra Savaglio[1,16], Patricia Schady[1], Sebastian Schmidl[5], Tassilo Schweyer[1,12], Vladimir Sudilovsky[1,17], Mohit Tanga[1], Adria C. Updike[18], Hendrik van Eerten[1], Karla Varela[1]

[1] Max-Planck Institut für Extraterrestrische Physik, Giessenbachstr. 1, 85748 Garching, Germany
[2] Excellence Cluster Universe, Technische Universität München, Boltzmannstr. 2, 85748, Garching, Germany
[3] Astrophysics Research Institute, Liverpool John Moores University, IC2, Liverpool Science Park, 146 Browlow Hill, Liverpool L3 5RF, United Kingdom
[4] Max-Planck Institut für Astrophysik, Karl-Schwarzschild-Str. 1, 85748 Garching, Germany
[5] Thüringer Landessternwarte Tautenburg, Sternwarte 5, 07778 Tautenburg, Germany
[6] European Southern Observatory, Alonso de Córdova 3107, Vitacura, Casilla 19001, Santiago 19, Chile
[7] INAF, Institute of Space Astrophysics and Cosmic Physics, via P. Gobetti 101, 40129 Bologna, Italy
[8] Scuola Normale Superiore, Piazza dei Cavalieri 7, I-56126 Pisa, Italy
[9] Departamento de Ciencias Fisicas, Universidad Andres Bello, Avda. Republica 252, Santiago, Chile
[10] European Southern Observatory, Karl-Schwarzschild-Str. 2, 85748 Garching, Germany
[11] American River College, Physics & Astronomy Dpt., 4700 College Oak Drive, Sacramento, CA 95841, USA
[12] Technische Universität München, Physik Dept., James-Franck-Str., 85748 Garching, Germany
[13] Institute of Experimental and Applied Physics, Czech Technical University in Prague, Horska 3a/22, 128 00 Prague 2, Czech Republic
[14] DARK Cosmology Center, Niels-Bohr-Institut, University of Copenhagen, Juliane Maries Vej 30, 2100 København, Denmark
[15] Department of Particle Physics & Astrophysics, Weizmann Institute of Science, Rehovot 76100, Israel
[16] Universita della Calabria, 87036 Arcavacata di Rende, via P. Bucci, Italy
[17] Astrophysics Data System, Harvard-Smithonian Center for Astrophysics, Garden St. 60, Cambridge, MA 02138, U.S.A.
[18] Roger Williams University, 1 Old Ferry Rd, Bristol, RI 02809, U.S.A.

**A new class of ultra-long duration (>10,000 s) gamma-ray bursts has recently been suggested[1,2,3]. They may originate in the explosion of stars with much larger radii than normal long gamma-ray bursts[3,4] or in the tidal disruptions of a star[3]. No clear supernova had yet been associated with an ultra-long gamma-ray burst. Here we report that a supernova (2011kl) was associated with the ultra-long duration burst 111209A, at $z$=0.677. This supernova is more than 3 times more luminous than type Ic supernovae associated with long gamma-ray bursts[5,6,7], and its spectrum is distinctly different. The continuum slope resembles those of super-luminous supernovae[8,9], but extends farther down into the rest-frame ultra-violet implying a low metal content. The light curve evolves much more rapidly than super-luminous supernovae. The combination of high luminosity and low metal-line opacity cannot be reconciled with typical type Ic supernovae, but can be reproduced by a model where extra energy is injected by a strongly magnetized neutron star (a magnetar), which has also been proposed as the explanation for super-luminous supernovae[20,20a].**

The gamma-ray burst (GRB) 111209A was detected by the *Swift* satellite at 07:12 UT on December 9, 2011. The X-ray and optical counterparts were discovered within minutes of the trigger[11]. The extraordinarily long duration of GRB 111209A was revealed by the continuous coverage with the Konus detector on the WIND spacecraft[12], extending from ~5 400 s before to ~10 000 s after the *Swift* trigger. The GRB occurred at a redshift of $z=0.677$, as determined from afterglow spectroscopy[3]. Its integrated equivalent isotropic energy output, $E_{iso} = (5.7\pm0.7)\times10^{53}$ erg [12], lies at the bright end of the distribution of long-duration GRBs.

The afterglow of GRB 111209A was observed over a period of about 70 days with the 7-channel optical/near-infrared imager GROND[13]. Starting around day 15, the optical light curve of the transient deviated from the earlier afterglow power law decay (Figure 1). The light curve remained essentially flat between day 15 and 30, and then started to decay again, approaching the host-galaxy level. After subtracting the afterglow and the well-modelled host galaxy emission (Methods, §1-§3), the excess emission is well constrained between rest-frame day 6 and 43 after the GRB (Figure 2). It is very similar in shape to other GRB-related supernovae (SN), but reaches a bolometric peak luminosity of $2.8^{+1.2}_{-1.0} \times 10^{43}$ erg/s (corresponding to $M_{bol}=-20.0$ mag) at 14 rest-frame days, a factor 3 higher than the brightest known GRB-SN (Figure 2).

A VLT/X-shooter spectrum was taken near the peak of the excess emission[3] (Dec. 29, 2011), 11.8 rest-frame days after the GRB. The afterglow and the (minimal) host contribution were subtracted (Methods, §3) and the resulting spectrum is shown in Figure 3 (blue line). The strong similarity of the evolution in time and colour to GRB-associated SNe, together with the spectral shape of the excess emission, leads us to conclude that this emission is caused by a supernova, designated SN 2011kl, associated with GRB 111209A.

Canonical long-duration GRBs are generally accepted to be linked to the core collapse of massive stars stripped of their outer H and He envelopes[5,6,7], since every spectroscopically confirmed supernova associated with a GRB has been a broad-lined SN Ic so far. Although the spectrum of SN 2011kl associated with the ultra-long GRB 111209A also shows no H or He, it is substantially different from classical GRB-SNe. It is surprisingly featureless redwards of 300 nm, lacking the undulations from spectral line blends typical of broad-lined SNe Ic associated with GRBs[5,6,7], and it does not drop in the 300–400 nm (rest-frame) region (Figure 3), suggesting a very low metal abundance. Applying standard parametrized SN light curve fits (Methods, §4), we derive an ejecta mass $M_{ej} = 3\pm1$ $M_\odot$ and a $^{56}$Ni mass of $1.0\pm0.1$ $M_\odot$, which implies a very high $^{56}$Ni/$M_{ej}$ ratio. This large $^{56}$Ni mass is not compatible with the spectrum, suggesting that $^{56}$Ni is not responsible for the luminosity, unlike canonical stripped-envelope SNe (Methods, §4).

Various models have been suggested to explain the ultra-long duration of GRB 111209A and other ULGRBs, but the otherwise inconspicuous spectral and timing properties of both the prompt and afterglow emission as well as the host properties provided no obvious clues[1-4,14-16]. With the detection of a supernova associated with the ultra-long GRB 111209A, we can immediately discard a tidal disruption interpretation[3]. Known supernovae from blue supergiants show hydrogen in their spectra and substantially different light curve properties[17], inconsistent with our observations, thus ruling out a blue supergiant progenitor[4]. Finally, additional emission from the interaction of the SN ejecta with circum-stellar material is unlikely as well (Methods, §5).

Our data suggest that SN 2011kl is intermediate between canonical overluminous GRB-SNe and super-luminous supernovae (SLSNe; Figure 3). SLSNe are a sub-class of SNe which are a factor ~100 brighter than normal core-collapse SNe, reaching $M_V \sim -21$ mag [8,9]. They show slow rise and late peak times (≈20–100 days as compared to typically 9–18 days). Their spectra are characterized by a blue continuum with a distinctive "W"-shaped spectral feature often interpreted as O II lines[8]. A spinning-down magnetic neutron star is the favoured explanation for the energy input powering the light curve[10]. The comparison of SN 2011kl with SLSNe is motivated by two observational facts: (1) the spectrum is a blue continuum, extending far into the rest-frame UV, and (2) the peak luminosity is intermediate between GRB-SNe and SLSNe. Our interpretation is motivated by the failure of both the collapsar and the standard fall-back accretion scenarios, because in these cases the engine quickly runs out of mass for any reasonable accretion rate and mass reservoir, and thus is unlikely to be able to power an ultra-long GRB.

We could reproduce the spectrum of SN 2011kl using a radiation transport code[18,19] and a density profile where $\rho \propto r^{-7}$, which is typical of the outer layers of SN explosions. The UV is significantly depressed relative to a blackbody, but much less depressed than in the spectra of GRB-SNe, indicating a lower metal content (consistent with 1/4 of the solar metallicity). The spectrum appears rather featureless owing to line blending. This follows from the high photospheric velocity, $v_{ph} \sim 20{,}000$ km/s (Figure 3). In contrast, SLSNe, which show more line features, have $v_{ph} \sim 10{,}000$ km/s. In the optical, on the other hand, only a few very weak absorption lines are visible in our SN spectrum. Our model only has $\sim 0.4 \, M_\odot$ of material above the photosphere. There is no evidence of freshly synthesized material mixed-in, unlike in GRB-SNe. This supports the notion that the SN light curve was not powered by $^{56}$Ni decay but rather by a magnetar.

The spectrum can be reproduced without invoking interaction, and the low metal abundance suggests that it is unlikely that much $^{56}$Ni was produced. We therefore consider magneto-rotational energy input as the source of luminosity. Using a simple formalism[20] describing rotational energy loss via magnetic dipole radiation, and relating the spin-down rate to the effective radiative diffusion time, one can infer the magnetar's initial spin period, $P_i$ and magnetic dipole field strength $B$ from the observed luminosity and time to light-curve peak, $t_{peak}$. The observed short $t_{peak}$ (~14 rest-frame days) and the moderate peak luminosity require a magnetar with initial spin period $P_i \sim 12$ ms for a magnetic field strength of $(6-9) \times 10^{14}$ G. Depending on the magnetic field, ejecta mass and kinetic energy are relatively uncertain ranging between 2 and 3 $M_\odot$ and $(2-9) \times 10^{51}$ erg, respectively (Methods, §6). These values are actually more typical of normal SNe Ib/c than of GRB-SNe, including SN 2006aj, the first SN identified as magnetar-powered[21]. The GRB energy can be reconciled with the maximum energy that can be extracted from a magnetar if the correction for collimation of the GRB jet is a factor of 1/50 or less, which is well within typical values for GRBs[22].

The idea of a magnetar as the inner engine powering GRB-SNe[23,24], SLSNe[10], or even events like Swift 1644+57 [25] (before consensus for this event favoured a relativistic tidal disruption), is not new. However, in all these cases the magnetar interpretation was one of several options providing reasonable fits to the data, but never cogent. Also, the suggestion that all GRB-SNe are magnetars[24] rather than collapsars, based on the clustering of the kinetic energy of the GRB-SNe near $10^{52}$ erg, the rotational power of a millisecond neutron star, was only circumstantial evidence. The supernova SN 2011kl is clearly different from canonical GRB-SNe, and requires (rather than only allows) a new explanation.

The ultra-long duration of the prompt emission of GRB 111209A and the unusual SN properties are probably related. We suggest that they are linked to the birth and subsequent action of a magnetar following the collapse of a massive star. The magnetar re-energizes the expanding ejecta and powers an over-luminous supernova. This particular SN 2011kl was not quite as luminous as typical SLSNe, and it may represent a population of events that is not easily discovered by SN searches but may have a relatively high rate. This scenario offers a link between GRB-SNe, ultra-long GRBs and SLSNe.

**Supplementary Information** is linked to the online version of that paper at www.nature.com/nature.

**Acknowledgements.** We are grateful to R. Lunnan and E. Berger for providing the spectrum of PS1-10bzj in digital form, and to A. Levan for the HST grism spectra of GRB 111209A. JG, ROD and DAK acknowledge support by the DFG cluster of excellence "Origin and Structure of the Universe" (www.universe-cluster.de). PS, JFG and MT acknowledge support through the Sofja Kovalevskaja award to P. Schady from the Alexander von Humboldt Foundation Germany. CD acknowledges support through EXTraS, funded from the European Union's Seventh Framework Programme for research, technological development and demonstration. SK, DAK and ANG acknowledge support by DFG. SSc acknowledges support by the Thüringer Ministerium für Bildung, Wissenschaft und Kultur. FOE acknowledges support from FONDECYT. ST is supported by DFG. Part of the funding for GROND (both hardware as well as personnel) was generously granted from the Leibniz-Prize to G. Hasinger. DARK is funded by the DNRF.

**Author contributions.** JG has led the observing campaign and the paper writing. DAK was responsible for the GROND data reduction, and performed the fitting of the afterglow light curve. FK derived the accurate GROND astrometry, PS the UVOT photometry, and AR the host fitting. PM suggested the magnetar interpretation and computed the spectral models. SP and CA performed the light-curve model fitting. FOE and EP assisted in spectral decomposition and the construction of the bolometric light curve. ST, SK, and GL provided crucial input and discussion. DAK, ANG, PMJA, JB, CD, JE, RF, JFG, SSc, TS, VS, MT, ACU and KV were performing the many epochs of GROND observations. TK, JPUF and GL provided and analysed the X-shooter spectrum. SSa, SK, RD, HvE have been instrumental in various aspects of the data interpretation.

**Author Information.** Reprints and permissions information is available at www.nature.com/reprints. The authors declare no competing financial interests. Correspondence and requests for materials should be addressed to jcg@mpe.mpg.de.

**Figure 1: Observed optical/near-infrared light curve following GRB 111209A.**
The light curve (GROND data: filled symbols; other data: open symbol) is the sum of the afterglow of GRB 111209A modelled by a broken power law (dashed line), the accompanying supernova 2011kl (thin red line) and the constant GRB/SN host galaxy emission (horizontal dotted line). All measurements (given with 1σ uncertainty) are relative to the Swift trigger time and as observed, apart from the Vega-to-AB transformation for the J-band. The solid violet line is the sum of afterglow and host in the u-band, with no sign of the supernova. The solid red line is the sum of afterglow, host and supernova for the r´-band. The vertical dotted line marks the time of the VLT/X-shooter spectrum.

**Figure 2: Light curve of the GRB 111209A supernova SN 2011kl.**
Bolometric light curve of SN 2011kl, corresponding to 230–800 nm rest frame (Methods, §1), compared with those of GRB 980425 / SN 1998bw [5], XRF 060218 / SN 2006aj [21], the standard type Ic SN 1994I [26], and the SLSNe PTF11rks [27] and PS1-10bzj [28] (among the fastest-declining SLSNe known so far), all integrated over the same wavelength band with 1σ error bars. Solid lines show the best-fitting synthetic light curves computed with a magnetar injection model [20] (dark blue; Methods, §6) and $^{56}$Ni powering (light blue; Methods, §4), respectively.

**Figure 3: Spectra comparison.**
The X-shooter spectrum of SN 2011kl, taken on Dec. 29, 2011 after GRB afterglow and host subtraction and moderate rebinning (Methods, §1; ED Fig. 2), with its flat shape and high UV flux is distinctly different from the hitherto brightest known GRB-SN 1998bw (red), but reminiscent of some SLSNe (top three curves)[28-30]. The three grey/black lines show synthetic spectra with different photospheric velocities (as labelled), demonstrating the minimum velocity required to broaden unseen absorption around 400 nm rest-frame (CaII, CII), but at the same time explain the sharp cut-off below 280 nm rest-frame. The y-scale is correct for SNe 2011kl and 1998bw; all other spectra are shifted for display purposes.

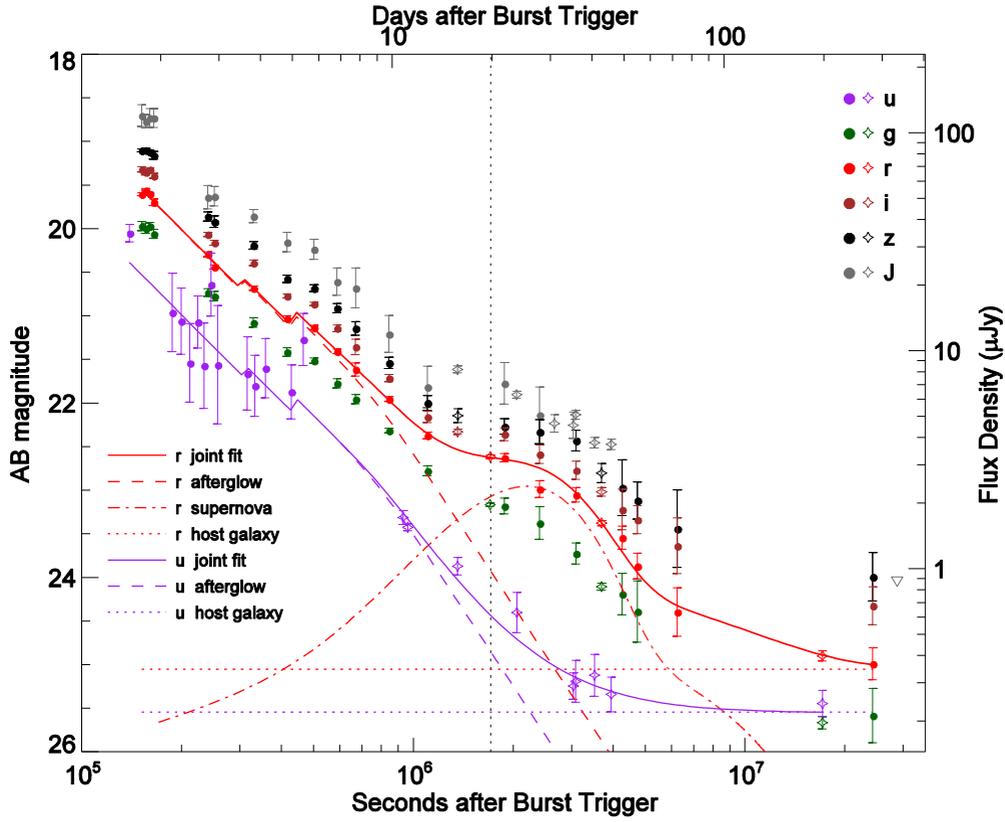

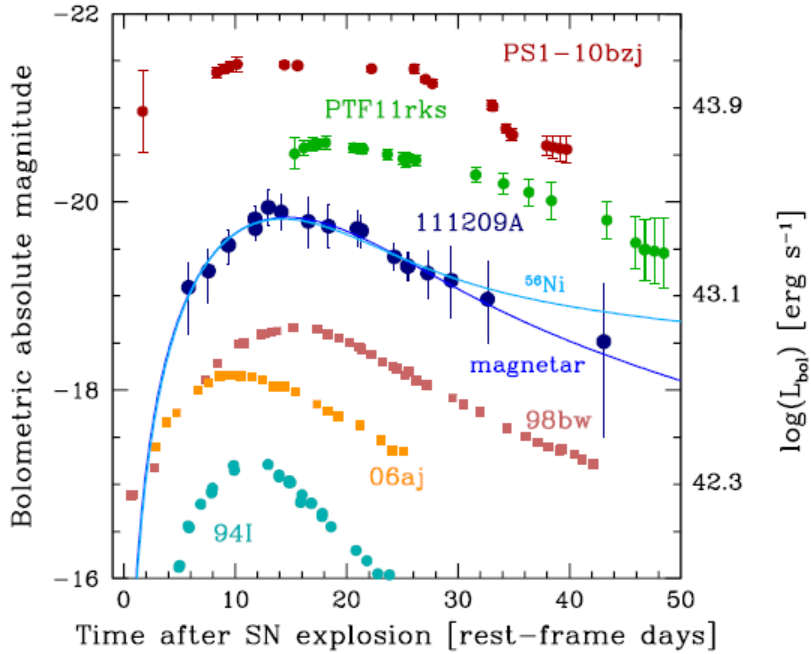

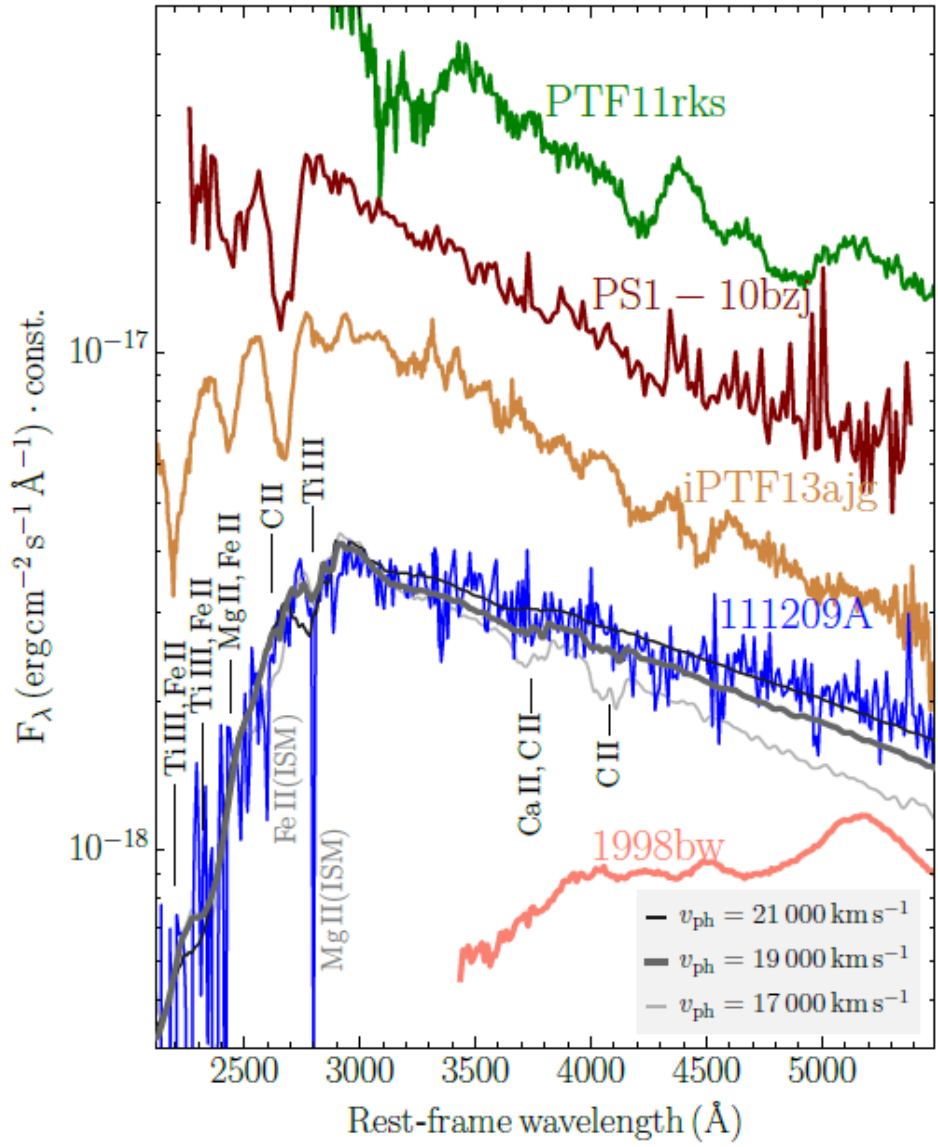

# Methods

## 1. Observations and Data Analysis

Simultaneous imaging in g'r'i'z'JHK$_s$ with the 7-channel imager GROND[13] was done on 16 epochs with logarithmic temporal spacing until 72 days after the GRB, when the nearby Sun prevented further observations, and a last epoch for host photometry was obtained 280 days after the GRB (Extended Data Table 1). GROND data have been reduced in the standard manner using pyraf/IRAF[31,32,33]. The optical imaging was calibrated against comparison stars obtained by observing a nearby SDSS field (immediately before the afterglow observation in the third night under photometric conditions) and calibrated against the primary SDSS[34] standard star network. The NIR data were calibrated against the 2MASS catalog. This results in typical absolute accuracies of ±0.03 mag in g´r´i´z´ and ±0.05 mag in JHK$_S$ (1σ errors are reported everywhere). All GROND measurements are listed in Extended Data Table 1, and the properties of the GRB afterglow proper, including the two kinks in the early afterglow light curve (Fig. 1) will be described in detail elsewhere [Kann, D.A. et al. 2015, Astron. Astrophys. (in prep.)].

We have made use of two other sources of measurements: First, we add u-band observations obtained with Swift/UVOT (Extended Data Table 2). UVOT photometry was carried out on pipeline-processed sky images downloaded from the Swift data centre[35] following the standard UVOT procedure[36], and is fully compatible with earlier, independent publications of the UVOT data[2,3]. Second, we add selected complementary data[3], in particular (i) HST F336W/F125W data from 11.1 and 35.1 days after the GRB, respectively; (ii) two epochs of VLT/FORS2 g´R$_C$i´z´ data during the SN phase, which agree excellently with our data due to their use of our GROND calibration stars; (iii) a late-time Gemini-S u´-band observation (198 days after the GRB).

With the constant host galaxy contribution accurately determined at late times in u´g´r´i´z´J (see §3 and Extended Data Fig. 4), the afterglow light curve shows clear evidence for a steeper afterglow decay at >10 days post-burst, particularly in the u´-band where there is essentially no contribution from the supernova (as evidenced by the spectrum) and which therefore can be used as a template for the pure afterglow contribution. We link the decay slopes for all filters to be the same. This provides the two decay slopes $\alpha_1$=1.55±0.01 and $\alpha_2$=2.33±0.06, with a break time of $t_b$=9.21±0.47 days. The u´-band fit is also shown in Figure 1 to visualize the decomposition. Apart from our much larger data set provided by our GROND observations, the difference to the decomposition of [3] is the fact that they ignored the host contribution in the redder bands at ≈30–50 days (though they actually note this in their work).

In order to create the SN light curve for each photometric band, we then subtracted both, the afterglow contribution in that band based on the extrapolation of the afterglow light curve, and the host galaxy contribution based on its spectral energy distribution; see §3. The error in the host galaxy subtraction is negligible as the host photometry is accurate to better than 10%, and the host contributes only between 5–15% to the total light during the SN bump. The error on the afterglow subtraction depends on whether or not the decay slope $\alpha_2$ remained constant after the last secure measurement right before the onset of the SN. The intrinsic GRB afterglow light curves at this late time are observed to only steepen, never flatten. Thus, our afterglow subtraction is conservative, and results in a lower limit for the SN luminosity.

The quasi-bolometric light curve of SN 2011kl was constructed from GROND g´r´i´z´J photometry and the supplementary data[3] as follows. First, the individual filter bands have been extinction-corrected with $A_V^{Gal}$ = 0.06 mag Galactic foreground[37], and rest-frame $A_V^{Host}$ = 0.12 mag as derived from the GRB afterglow spectral energy distribution fitting. By deriving quadratic polynomials for sets of three consecutive filters (Simpson's rule) they were then combined to create a quasi-bolometric light curve.

The quadratic polynomials are then integrated over rest-frame wavelength from 3860/(1+z) Å (blue edge of the g´-band filter) to 13560/(1+z) Å (red edge of the J filter). The k-correction was computed from the spectral energy distribution. In order to transform the integrated flux into luminosity, we employed a luminosity distance of d=4080 Mpc, using concordance cosmology ($\Omega_\Lambda$=0.73, $\Omega_m$=0.27, and $H_0$=71 km s$^{-1}$ Mpc$^{-1}$).

No correction for the contribution of the unobserved near-infrared part of the spectrum has been applied to both, SN 2011kl and SN 1998bw (Figure 2), because this emission is usually sparsely sampled in wavelength and time, and thus is largely based on assumptions (and no data is available for the plotted SLSNe). For SN 2011kl we lack any rest-frame near-infrared measurements. We acknowledge that therefore the bolometric luminosity might be underestimated by 5%–30%. Other than that, all bolometric light curves shown in Fig. 2 are integrated over the same wavelength band (except for the UV band, which contributes less than a few percent at and after maximum). The SLSN light curves are plotted according to the observational constraints of their maxima, i.e. g-band peak at 16.8 days rest-frame for PTF11rks [27] and using the first measurement 17.5 days before maximum as lower limit for PS1-10bzj [28].

The VLT/X-shooter[38] spectrum, taken on Dec. 29, 2011 (19.8 days after the GRB, 11.8 rest-frame days, and 2 days prior to the SN maximum), has been reduced with the ESO X-shooter pipeline v2.2.0, in particular for flat-fielding, order tracing, rectification and initial wavelength calibration with an arc lamp. During rectification, a dispersion of 0.4Å/pixel has been used in the UVB/VIS arm, minimizing correlated noise but maintaining sufficient spectral resolution for resolving lines down to ≈50 km/s, i.e. a velocity dispersion of 20 km/s. Our own software is used for bad-pixel and cosmic-ray rejection, as well as sky-subtraction and frame shifting and adding[39]. Optimal extraction is applied to the resulting 2D frames, and the one-dimensional spectrum is finally flux calibrated separately for each arm against the GROND photometry. Spectral binning has no effect on the steepness of the slope (Extended Data Figure 1). The NIR arm does not contain any useful signal, as do the two HST grism spectra[3] (Extended Data Figure 2).

The observed spectrum is the sum of light from the GRB afterglow, the GRB host galaxy, and the supernova SN 2011kl. After correcting for $A_V^{Gal}$ = 0.06 mag Galactic foreground[37] extinction, we corrected for the contribution of the host galaxy using a template fit (§3) on the host photometry (including the J-band measurement of [3]), and subtracted the afterglow based on the extrapolation of the g´r´i´z´ GROND light curves to the time of the X-shooter observation. After conversion to the rest-frame, we corrected for intrinsic reddening of E(B−V)=0.04±0.01 mag derived from the GROND afterglow SED fitting (see Extended Data Figure 3 for the effect of each of these steps).

## 2. Association of GRB afterglow, supernova, and host galaxy

We detect narrow absorption lines of Mg II($\lambda$2796, $\lambda$2803), Mg I($\lambda$2852) and Fe II($\lambda$2344, $\lambda$2374, $\lambda$2382, $\lambda$2586, $\lambda$2600) in the SN 2011kl spectrum. No change in equivalent widths and redshift is apparent when compared to the afterglow spectrum[3,39] taken 0.75 days after the GRB. Moreover, these equivalent widths are typical of those seen from host galaxies of bright long-duration GRBs. This relates the SN to the same host galaxy as GRB 111209A.

No offset is measurable in GROND images between GRB afterglow and supernova ($\delta$RA<0."032, $\delta$Decl<0."019), which implies that the two events are co-spatial within <200 pc.

## 3. The host galaxy

During the late-epoch GROND observation the host galaxy is clearly detected in g´r´i´z´ in the 3–5$\sigma$ range (last entry in Extended Data Table 1). We add HST F336W and Gemini u'J from [3]. Noting that the supernova does not contribute significantly any more during these late epochs (with expected AB magnitudes g'≈28.5, r'≈28.0, i'≈27.5, z'≈27.2 mag), we employ `LePHARE`[40] and use the best-fit model (a low-mass, star-forming galaxy) as a template for the host subtraction (see Extended Data Figure 3 and 4). Inferences on the physical properties of the host from this fitting will be published elsewhere [Kann, D.A. et al. 2015, Astron. Astrophys. (in prep.)] and absorption/emission line information from the optical/NIR X-shooter spectra are given in [39]. We note though that the low metal content seen in the SN spectrum is in accord with the very low host galaxy metallicity (10%-40%), which is somewhat unusual for such a low-redshift object but commonly seen in SLSN hosts.

## 4. Radioactivity cannot power the supernova peak

Modelling the bolometric light curve according to the standard scheme of $^{56}$Ni powering[41] and augmented by Co decay[42], an ejecta mass of 3.2±0.5 M$_\odot$ and a $^{56}$Ni mass of 1.0±0.1 M$_\odot$ is derived (we used $v_{ph}$ = 20,000 km/s, and a grey opacity of 0.07±0.01 cm$^2$ g$^{-1}$, constant in time). The derived $^{56}$Ni mass is anomalously large for SNe Ib/c, including GRB-SNe[43]. Such a large $^{56}$Ni mass is difficult to reconcile with the very low opacity in the blue part of the spectrum. The continuum flux keeps rising down to 300 nm rest-frame without any sign of suppression implying very low metal line opacity. Also, the ejected mass of ≈3 M$_\odot$ as deduced from the light curve width does not resonate with the large $^{56}$Ni mass.

While it has been suggested that part of the $^{56}$Ni could be synthesised in the accretion disk[44], this is unlikely to proceed at a rate needed in our case. Recent numerical simulations show that for a wide range of progenitor masses (13–40 M$_\odot$), initial surface rotational velocities, metallicities and explosion energies, the required disk mass of more than 1 M$_\odot$ (corresponding to ~0.2 M$_\odot$ $^{56}$Ni) is difficult to achieve[45], for both cases of compact objects: (i) in the case of heavy fallback, leading to the collapse of the central object into a black hole, the explosion energy is required to be small (few × 10$^{48}$ erg), and more importantly, the disk forms only after a few months due to the large fallback time (~10$^6$ s). (ii) in the case of little fallback, leaving a neutron star behind, only fine-tuned conditions produce fallback disks at all, and these then have lifetimes of at most several hundred seconds.

Thus, a different mechanism must power the SN light curve during the first ~40 days (rest frame).

## 5. Enhanced emission due to interaction with the circumburst medium?

Given the large luminosity, we considered additional emission from the interaction of the supernova ejecta with the circumstellar medium as an alternative possibility. In that case, one may expect narrow Balmer emission lines. While we detect very narrow ($\sigma$=35 km/s) H$\alpha$, H$\beta$ and [OIII] lines in emission, the Balmer fluxes are compatible with the forbidden line flux and with an origin from the global low (0.02 M$_\odot$/yr) star formation rate in this low-metallicity (10%-40% solar) host galaxy[39]. On the other hand, if the progenitor star was heavily stripped, no circumstellar H may be present. Another, more serious constraint is the very blue SN spectrum, which would require a very low density to minimize extinction (though dust may be destroyed by the initial GRB and SN light). This may be at odds with the requirement that the density is high enough to generate the few $10^{43}$ erg s$^{-1}$ of radiative luminosity observed around the peak.

## 6. Modelling

We have been able to reproduce the spectrum of SN 2011kl using a radiation transport code[18,19] and a density profile where $\rho \propto r^{-7}$, which is typical of the outer layers of SN explosions. The spectra appear rather featureless but this does not mean that there is no absorption: the UV is significantly depressed relative to a blackbody. However, it is much less depressed than the spectra of GRB-SNe, indicating a lower metal content. Many metal lines are active in the UV (Fe, Co, Ti, Cr). The smooth appearance of the UV spectrum is the result of the blending of hundreds of lines caused by the large range of wavelengths over which lines are active (line blanketing). Indeed, the photospheric velocity (and density) determines the degree of line blending. We used here photospheric velocities of $v_{ph} \sim 20,000$ km/s (grey/black lines in Figure 3), and can see increasingly featureless spectra as $v_{ph}$ increases and lines are active at higher velocities (larger blueshift), demonstrating the minimum velocity required to broaden unseen absorption around 400 nm rest-frame (CaII, CII), but at the same time explain the sharp cut-off below 280 nm rest-frame. The strongest lines that shape this strong blue cut-off are labelled in black (grey 'ISM' label mark MgII/FeII absorption lines in the host galaxy). Most of these are blended and do not stand out as individual features, unlike in classical SLSNe which have $v_{ph} \sim 10,000$ km/s. In the optical, on the other hand, only few very weak absorption lines are visible in our SN spectrum. These are due to Ca II and C II lines. O II lines are not detected, and would require large departures from thermal equilibrium because of the very high ionization/excitation potential of their lower levels (20-30 eV). This suggests the presence of X-rays in SLSNe, probably produced by shocks. Our model only has $\sim 0.4$ M$_\odot$ of material above the photosphere. The metal content is quite low. It is consistent with 1/4 of the solar metallicity, which could be the metallicity of the star whose explosion caused the GRB and the SN, and there is no evidence of freshly synthesised material mixed-in, unlike in GRB-SNe. This supports the notion that the SN light curve was not powered by $^{56}$Ni decay but rather by a magnetar. Figure 3 shows this model with three different photospheric velocities overplotted on the X-shooter spectrum.

The spectrum can be reproduced without invoking interaction, but the metal abundance is so low that it is unlikely that much $^{56}$Ni has been produced. We therefore consider magneto-rotational energy input as the source of luminosity. Depending on the relative strength of magnetar and radioactive decay energy deposition, different peak luminosities as well as rise and decay times can be obtained[20]. One particularly pleasant feature of the magnetar mechanism is that it does not necessarily suffer from strong line blanketing, thus providing a more natural explanation for the observed spectrum.

Using a simple formalism describing rotational energy loss via magnetic dipole radiation and relating the spin-down rate to the effective radiative diffusion time, one can infer the magnetar's initial spin period $P_i$ and magnetic dipole field strength from the observed luminosity and time to light curve peak $t_{peak}$. One million combinations of the parameters $P_i$, $B$, $M_{ej}$ and $E_K$ were sampled and ranked according to the goodness of fit relative to the data. All best solutions cluster at $P=12.2\pm0.3$ ms and have $B=7.5\pm1.5\times10^{14}$ G, required by the observed short $t_{peak}$ (~14 rest-frame days) and the moderate (for a magnetar) peak luminosity. The mass and energy of the ejecta are less well determined, as they depend on the energy injection by the magnetar, and also due to the unknown distribution of mass in velocity space below the photosphere. We find a rather low ejected mass $M_{ej}=2.4\pm0.7$ M$_\odot$, and energy $E_K=(5.5\pm3.3)\times10^{51}$ erg. Different photospheric velocities of e.g. 10,000, 15,000 and 20,000 km/s lead to different ejecta masses of 1.1, 1.7 and 2.2 M$_\odot$, but produce indistinguishable light curves with $M_{Ni}=1.0\pm0.1$ M$_\odot$. Note though that not every combination of $P_i$, $M_{ej}$ and $E_K$ yields similar results. The GRB energy can be reconciled with the maximum energy that can be extracted from a magnetar if the correction for collimation of the GRB jet is a factor of 1/50 or less, which is well within typical values for GRBs[22].

## 7. Code availability

The code used in [18, 19] is available on request from mazzali@mpa-garching.mpg.de.

## 8. Supplementary References

**Extended Data**

**Extended Data Figure 1 | Binning has no effect on spectral slope.**
Original X-shooter spectrum in the UVB (top) and VIS (bottom) arms shown in gray (0.4 Å/pix; prior to host and afterglow subtraction), with the re-binned (factor 20) spectrum overplotted in black. The binning does not change the steepness of the spectrum, in particular not at the blue end.

**Extended Data Figure 2 | Long-wavelength spectra.**
Full X-shooter spectrum near maximum light of SN 2011kl, as well as two HST grism spectra taken one week before and after the supernova maximum, respectively (both taken from [3]. Above 500 nm rest-frame, none contains any informative absorption lines (all absorption structures seen are from the Earth atmosphere).

**Extended Data Figure 3 | Step-by-step corrections of the supernova spectrum.**
Sequence of analysis steps for the X-shooter spectrum, from the only galactic foreground corrected observed spectrum (top/very light blue), over host subtraction (light blue) and afterglow+host subtraction (blue). The break at 500 nm observer-frame (300 nm rest-frame) and the steep slope towards the UV are inherent to the raw spectrum, not a result of afterglow or host subtraction.

**Extended Data Figure 4 | Observed spectral energy distribution of the host galaxy of GRB 111209A.**
Plotted in blue are GROND g´r´i´z´ detections with 1σ errors (crosses) and GROND JHK$_S$ upper limits (3σ; triangles) of the host galaxy of GRB 111209A. Data taken from [3] are F336W (green), Gemini g´r´ detections (red crosses) and the J-band upper limit (red triangle). The best-fit LePHARE template of a low-mass, low-extinction, young star-forming galaxy is shown which is very typical for GRB host galaxies.

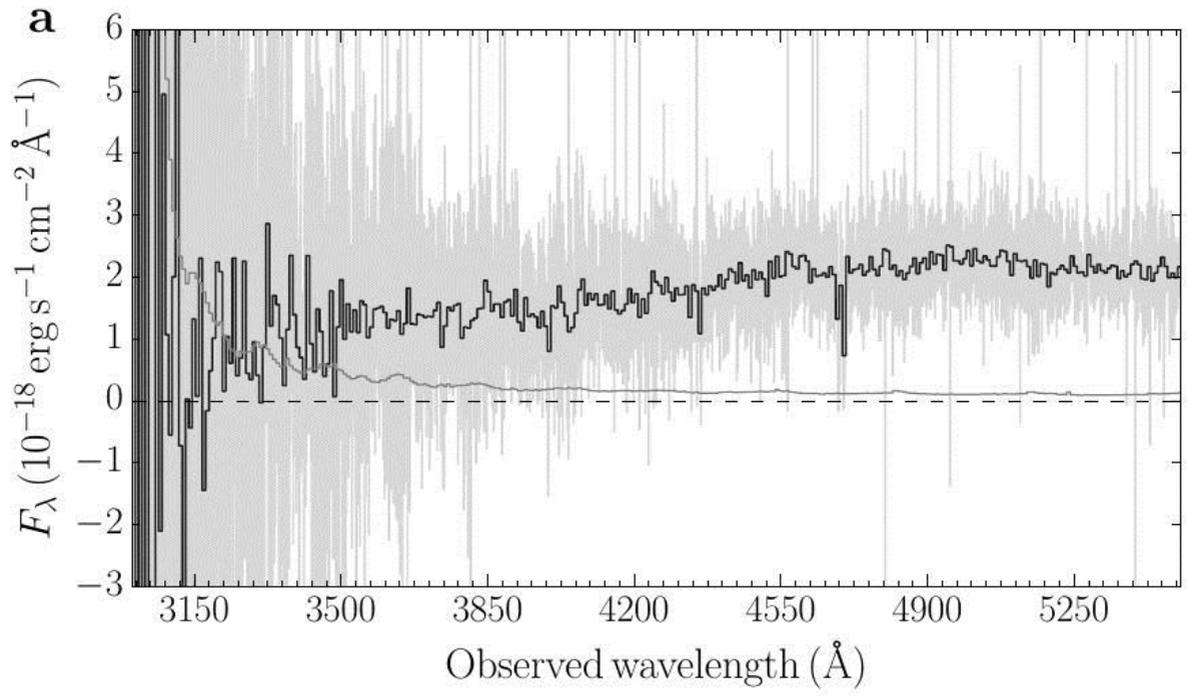
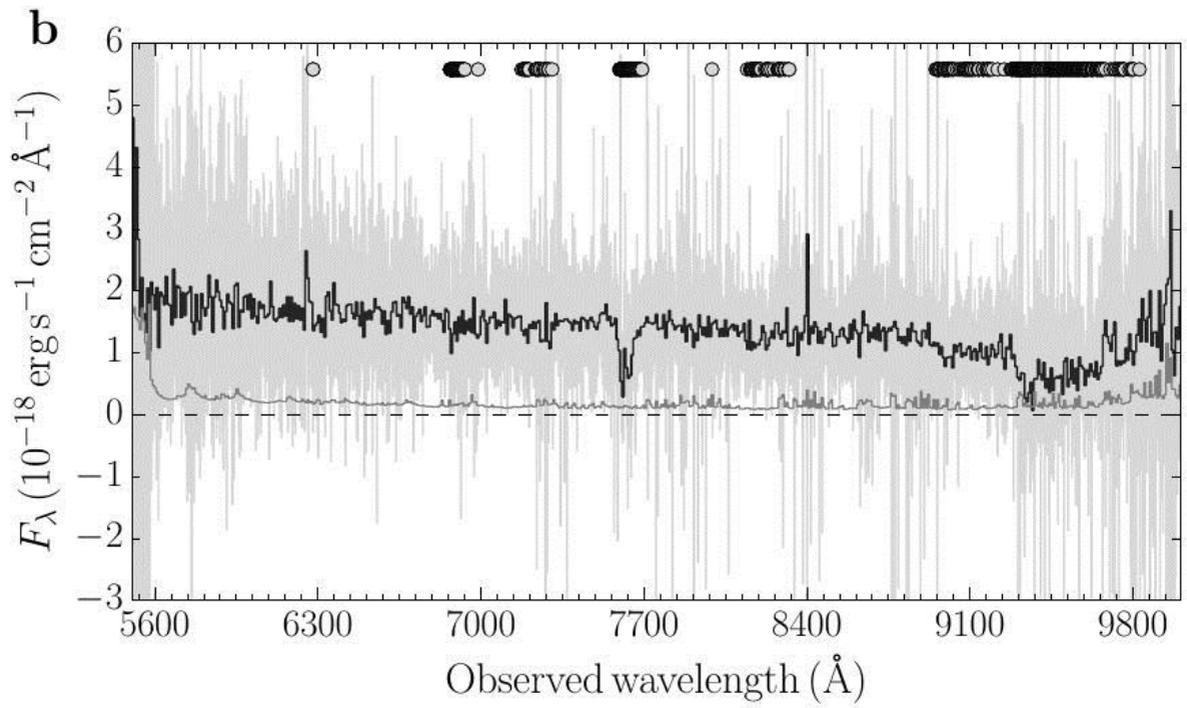

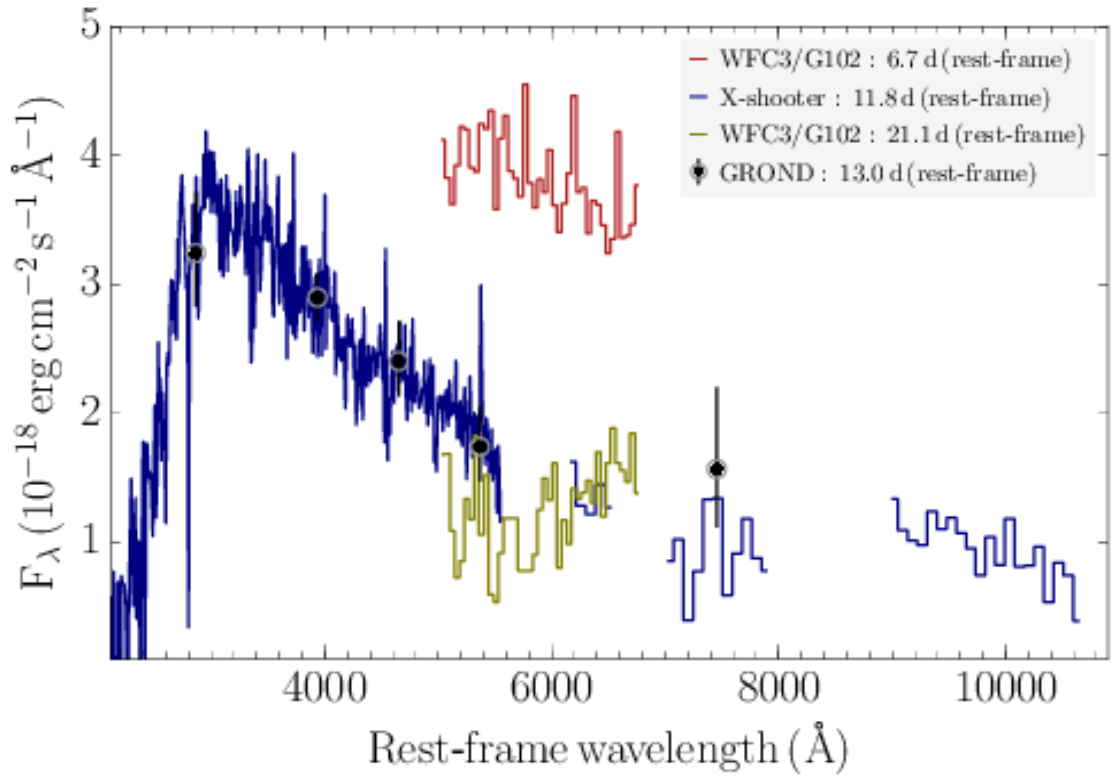

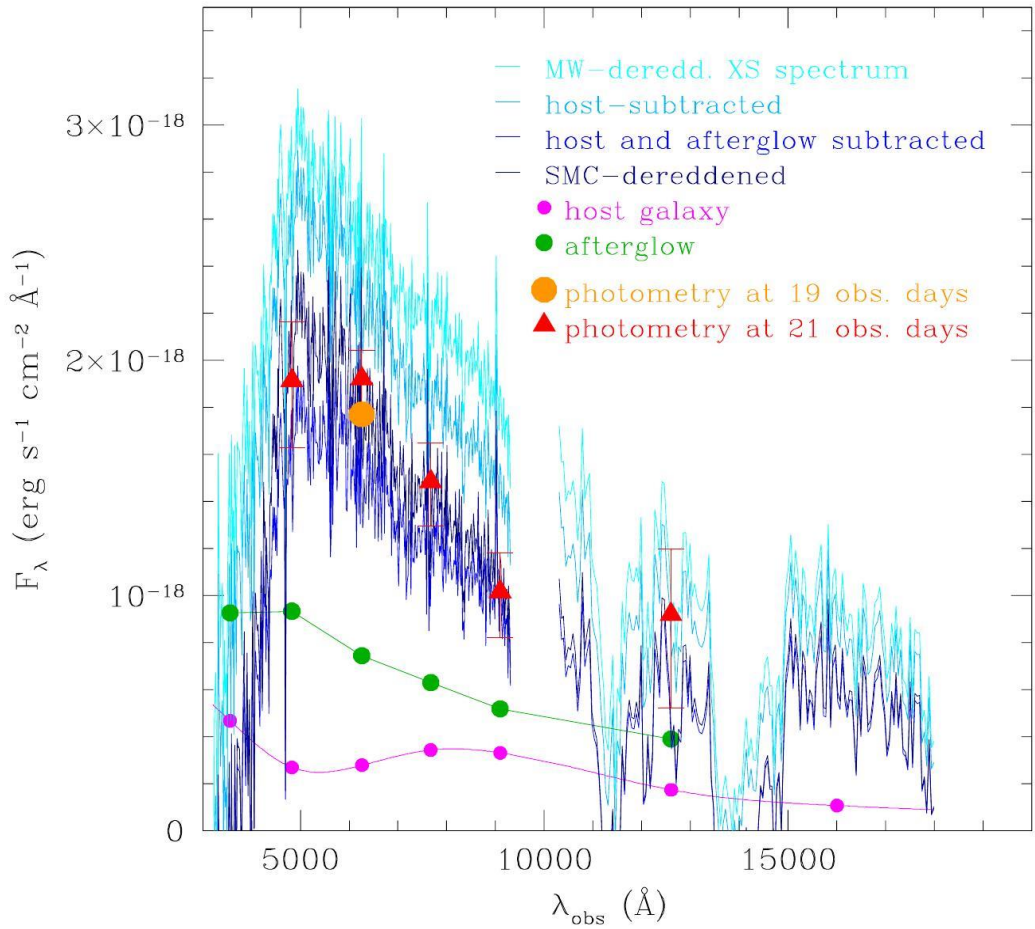

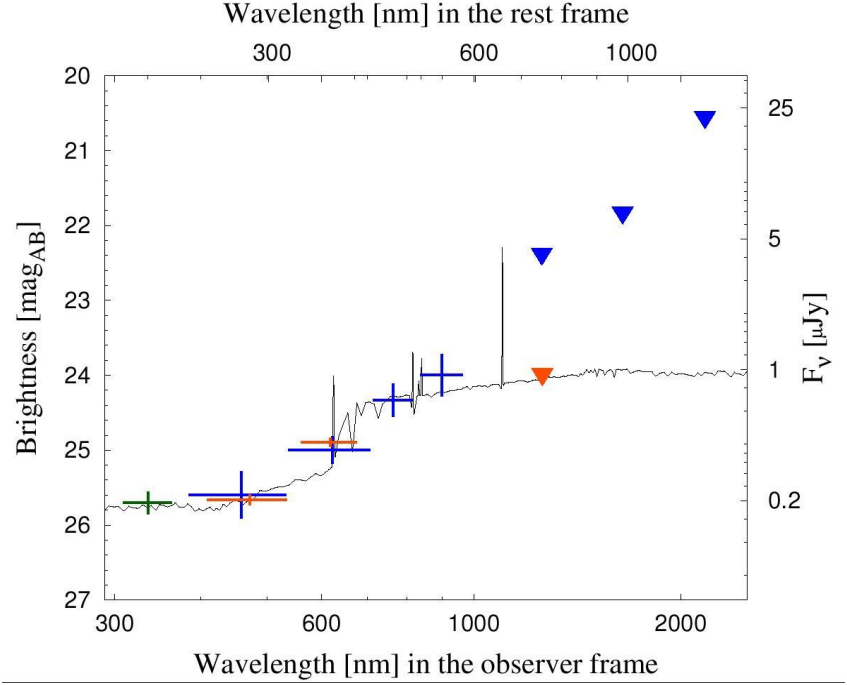

**Extended Data Table 1 | GROND observations of the afterglow, supernova and host of GRB 111209A.**

The Δt time gives the mid-time of the observation relative to the Swift trigger time. All magnitudes are in the AB system and not corrected for Galactic foreground extinction. Conversion to Vega magnitudes: $g'_{AB} - g'_{Vega} = -0.062$ mag, $r'_{AB} - r'_{Vega} = 0.178$ mag, $i'_{AB} - i'_{Vega} = 0.410$ mag, $i'_{AB} - i'_{Vega} = 0.543$ mag, $J_{AB} - J_{Vega} = 0.929$ mag, $H_{AB} - H_{Vega} = 1.394$ mag, $K_{S,AB} - K_{S,Vega} = 1.859$ mag. Corrections for Galactic extinction are $A_g = 0.066$ mag, $A_r = 0.046$ mag, $A_i = 0.034$ mag, $A_z = 0.025$ mag, $A_J = 0.015$ mag, $A_H = 0.010$ mag, $A_{K_S} = 0.006$ mag.

| Δt (ks) | $T_{VIS}$ (s) | g' (mag) | r' (mag) | i' (mag) | z' (mag) | $T_{NIR}$ (s) | J (mag) | H (mag) | $K_s$ (mag) |
|---|---|---|---|---|---|---|---|---|---|
| 151.49 | 460 | 20.05±0.05 | 19.66±0.02 | 19.36±0.03 | 19.13±0.02 | 480 | 18.72±0.12 | 18.31±0.12 | 17.84±0.15 |
| 155.91 | 460 | 20.07±0.06 | 19.62±0.02 | 19.39±0.03 | 19.13±0.02 | 480 | 18.79±0.08 | 18.31±0.11 | 17.87±0.15 |
| 160.33 | 460 | 20.05±0.04 | 19.65±0.02 | 19.36±0.02 | 19.15±0.02 | 480 | 18.75±0.10 | 18.35±0.10 | 18.01±0.16 |
| 164.70 | 460 | 20.14±0.05 | 19.75±0.04 | 19.43±0.03 | 19.19±0.04 | 480 | 18.75±0.11 | 18.40±0.12 | 18.09±0.18 |
| 239.81 | 919 | 20.81±0.04 | 20.35±0.03 | 20.11±0.02 | 19.89±0.05 | 960 | 19.66±0.13 | 19.01±0.14 | 18.71±0.17 |
| 250.95 | 919 | 20.85±0.06 | 20.49±0.02 | 20.20±0.03 | 19.95±0.07 | 960 | 19.65±0.11 | 19.11±0.12 | 18.98±0.21 |
| 329.17 | 1133 | 21.16±0.06 | 20.74±0.03 | 20.43±0.03 | 20.22±0.05 | 1920 | 19.87±0.08 | 19.39±0.12 | 19.10±0.18 |
| 415.47 | 1838 | 21.49±0.05 | 21.08±0.03 | 20.81±0.02 | 20.60±0.04 | 1920 | 20.17±0.11 | 19.88±0.16 | 19.65±0.26 |
| 501.08 | 1838 | 21.59±0.05 | 21.19±0.02 | 20.90±0.02 | 20.71±0.05 | 1920 | 20.25±0.11 | 19.99±0.15 | 19.94±0.32 |
| 588.10 | 1838 | 21.85±0.05 | 21.46±0.03 | 21.18±0.04 | 20.94±0.05 | 1920 | 20.62±0.16 | 20.25±0.19 | 19.67±0.27 |
| 669.18 | 919 | 22.03±0.05 | 21.67±0.08 | 21.40±0.09 | 21.17±0.08 | 960 | 20.70±0.23 | 20.36±0.29 | 19.86±0.35 |
| 843.66 | 1379 | 22.39±0.03 | 22.01±0.03 | 21.75±0.04 | 21.57±0.06 | 1440 | 21.23±0.21 | 20.71±0.40 | 20.49±0.46 |
| 1101.93 | 2420 | 22.86±0.06 | 22.42±0.04 | 22.20±0.07 | 22.03±0.09 | 2160 | 21.83±0.24 | 20.82±0.25 | 20.57±0.52 |
| 1880.55 | 2952 | 23.26±0.09 | 22.68±0.05 | 22.40±0.07 | 22.30±0.09 | 2400 | 21.79±0.24 | 21.76±0.27 | 20.70±0.75 |
| 2401.32 | 4502 | 23.45±0.19 | 23.00±0.09 | 22.63±0.11 | 22.36±0.14 | 3600 | 22.15±0.32 | 21.86±0.36 | >20.32 |
| 3090.97 | 3630 | 23.80±0.12 | 23.11±0.08 | 22.81±0.10 | 22.46±0.12 | 3240 | >22.25 | >21.85 | >20.22 |
| 4258.44 | 5384 | 24.27±0.24 | 23.60±0.13 | 23.26±0.23 | 23.00±0.32 | 4560 | >21.54 | >21.05 | >19.19 |
| 4732.20 | 5422 | 24.47±0.35 | 23.92±0.15 | 23.38±0.16 | 23.15±0.21 | 4560 | >22.06 | >21.62 | >20.33 |
| 6241.88 | 2758 | >24.57 | 24.45±0.28 | 23.68±0.32 | 23.47±0.44 | 2880 | >21.52 | >20.91 | >20.06 |
| 24277.46 | 3752 | 25.66±0.31 | 25.04±0.18 | 24.36±0.22 | 24.02±0.28 | 3600 | >22.39 | >21.84 | >20.56 |

**Extended Data Table 2 | UVOT observations of the afterglow of GRB 111209A.**

The Δt time gives the mid-time of the observation relative to the *Swift* trigger time, and all magnitudes are in the AB system and not corrected for Galactic foreground extinction. Conversion to Vega magnitudes: $u_{AB} - u_{Vega} = 1.02$ mag (as given at http://swift.gsfc.nasa.gov/analysis/uvot_digest/zeropts.html). The correction for Galactic extinction, using $E_{(B-V)} = 0.017$ mag [36] and the Galactic extinction curve [46] is $A_u = 0.085$ mag.

| Δt (ks) | T (s) | u (mag) |
|---|---|---|
| 139.3566 | 546.0 | $20.23^{+0.11}_{-0.10}$ |
| 187.4401 | 157.0 | $21.14^{+0.77}_{-0.45}$ |
| 199.3795 | 157.0 | $21.24^{+0.58}_{-0.38}$ |
| 211.8172 | 157.0 | $21.72^{+0.77}_{-0.45}$ |
| 223.9091 | 235.5 | $21.25^{+0.42}_{-0.30}$ |
| 233.6637 | 235.5 | $21.75^{+0.90}_{-0.49}$ |
| 245.1895 | 156.9 | $20.82^{+0.55}_{-0.36}$ |
| 256.7393 | 157.0 | $21.74^{+2.17}_{-0.68}$ |
| 286.4793 | 84.7 | >20.66 |
| 315.6230 | 314.1 | $21.84^{+0.70}_{-0.42}$ |
| 332.6649 | 382.4 | $21.98^{+0.52}_{-0.35}$ |
| 357.8214 | 844.0 | $21.78^{+0.51}_{-0.34}$ |
| 428.4023 | 578.3 | $22.05^{+0.44}_{-0.31}$ |
| 465.3887 | 342.0 | $21.45^{+0.42}_{-0.30}$ |